\documentclass[aip,amsmath,amssymb]{revtex4-1}

\draft 

\usepackage{graphicx}
\usepackage{dcolumn}
\usepackage{bm}

\usepackage[utf8]{inputenc}
\usepackage[T1]{fontenc}
\usepackage{mathptmx}
\usepackage{etoolbox}

\usepackage[caption=false]{subfig}
\usepackage{booktabs}
\usepackage[export]{adjustbox}
\usepackage{multirow}
\usepackage{rotating}
\usepackage{tablefootnote}
\usepackage{mathtools}

\makeatletter
\def\@email#1#2{%
 \endgroup
 \patchcmd{\titleblock@produce}
  {\frontmatter@RRAPformat}
  {\frontmatter@RRAPformat{\produce@RRAP{*#1\href{mailto:#2}{#2}}}\frontmatter@RRAPformat}
  {}{}
}%
\makeatother

\begin{document}
\preprint{AIP/123-QED}

\title{Effect of Antral Motility on Food Hydrolysis and Gastric Emptying from the Stomach: Insights from Computational Models} 



\author{Sharun Kuhar}
    \affiliation{Department of Mechanical Engineering, Johns Hopkins University, Baltimore, MD, USA}
\author{Jae Ho Lee}
    \altaffiliation{Present address: Center for Drug Evaluation and Research, U.S. Food and Drug Administration, Silver Spring, MD, USA.}
    \affiliation{Department of Mechanical Engineering, Johns Hopkins University, Baltimore, MD, USA}
\author{Jung-Hee Seo}
    \affiliation{Department of Mechanical Engineering, Johns Hopkins University, Baltimore, MD, USA}
\author{Pankaj J Pasricha}
    \affiliation{Division of Gastroenterology and Hepatology, Johns Hopkins School of Medicine, Baltimore, MD, USA}
\author{Rajat Mittal}
    \email{mittal@jhu.edu}
    \affiliation{Department of Mechanical Engineering, Johns Hopkins University, Baltimore, MD, USA}
    \affiliation{Institute for Computational Medicine, Johns Hopkins University, Baltimore, MD, USA}
    \affiliation{School of Medicine, Johns Hopkins University, Baltimore, MD, USA}
    

\date{\today}

\begin{abstract}
The peristaltic motion of the stomach walls combines with the secretion of enzymes to initiate the process that breaks down food. Computational modelling of this phenomenon can help reveal the details that would be hard to capture via \emph{in-vivo} or \emph{in-vitro} means. In this study, the digestion of a liquid meal containing protein is simulated in a human-stomach model based on imaging data. Pepsin, the gastric enzyme for protein hydrolysis, is secreted from the proximal region of the stomach walls and allowed to react with the contents of the stomach. The jet velocities, the emptying rate, and the extent of hydrolysis are quantified for a control case, and also for three other cases of reduced motility with varying peristaltic amplitudes. The findings quantify the effect of motility on the rate of food breakdown and emptying, and correlate the observations with the mixing in the stomach induced by the antral contraction waves.

\end{abstract}


\maketitle 



\section{Introduction}


The human stomach stores, mixes, processes, and grinds ingested food.
Understanding these
functions is becoming increasingly important as issues related to  digestion related disorders and nutrition are becoming prevalent  \cite{barnard_trends_2010,casini_trends_2015,oberlander_globalisation_2017}. 
For researchers in the field of medicine, stomach motility plays
a key role in gastric disorders such as gastroesophageal reflux disease (GERD), gastroparesis (delayed emptying), 
dumping syndrome (rapid emptying), as well as in the pathophysiology of diabetes and 
functional dyspepsia\cite{grover_gastroparesis_2019,koch_diabetic_2015,ukleja_dumping_2005,mizuta_recent_2006}, because
the rate at which the stomach empties its contents affects the rate at which the
nutrients get absorbed in the intestines. Gastric emptying is also important for drug delivery via oral administration 
as it determines the time it takes for effects of the drug to manifest themselves \cite{lee_computational_2022}.
For researchers in the arena of nutrition and food design, a better understanding of the biomechanical processes involved in gastric digestion could help
answer questions related to bioavailability of nutrients and the variation in emptying rates and food hydrolysis with food properties. 

    
The physical and chemical action of the stomach is highly regional. The stomach has three major regions - the fundus, the corpus 
and the antrum. The fundus and corpus grow or shrink to accommodate the
increasing volume of ingested food without increasing the pressure in the lumen. They also contain glands that
secrete acid and enzymes to breakdown food \cite{bornhorst_gastric_2017}. The gastric pacemaker is also present along the greater curvature in the corpus. The pacemaker is responsible for initiating  periodic peristaltic waves, known as antral contraction waves (ACWs), that travel aborally along the stomach walls.
The amplitude of these waves is small in the region of inception but it
rises gradually till the waves reach the antrum \cite{pal_gastric_2004,schulze_imaging_2006}. At the distal end of the 
antrum is the pylorus - a small orifice connecting the stomach to the duodenum. The orifice is surrounded by muscular
tissues responsible for opening and closing the pylorus. Right before the ACWs terminate at
the pylorus, a segmental contraction leads to a high-amplitude collapse of the walls; this event is known as the
terminal antral contraction (TAC). TAC propels the trapped gastric contents away from the 
collapsing orifice and the coordination between the opening/closing of pylorus and the TAC is crucial to 
the dynamics of gastric emptying\cite{ishida_quantification_2019}. 

    
These muscular activities vary from meal-to-meal and person-to-person. Liquid meals are emptied much faster
than solid meals \cite{kelly_gastric_1980}. Lower viscosity meals generate deeper indentation ACWs \cite{koziolek_simulating_2013} than higher viscosity ones.
There also exists a feedback loop between the duodenum and the antrum that regulates the rate of caloric availability in the duodenum to about 2-4 kcal/min \cite{brener_regulation_1983}. The meal structure and softness plays a role in the emptying rate as well because the 
pyloric orifice ensures that only particles smaller than 2 mm are emptied while
the rest continue to undergo fragmentation and dissolution within the antrum by the action of the retropulsive jet and the ACWs \cite{schulze_imaging_2006}. 
Thus, gastric mixing and emptying are complex processes with intricate neurohormonal control mechanisms that adjust the motility based on multiple food properties.

    
Studying the gastric chemo-fluid dynamics via experiments poses several challenges. \emph{In-vivo} approaches
are cost-intensive and ethically constrained. 
\emph{In-vitro} models, on the other hand, fail to replicate the complex wall motion as well as the chemical response to 
the food \cite{bornhorst_gastric_2014}. However, the stomach geometry and motility information gleaned from MRI scans 
can be used to construct computational (in-silico) models that overcome many of the limitations of these other approaches. This has become possible only
in the recent decades with the advancement in computational power and algorithms. 


There are relatively few examples of 
studies that have tried to investigate gastric mixing and emptying via computational models. The first work that used computational fluid dynamics (CFD) to investigate flow inside the 
stomach was based on a 2D stomach model \cite{pal_gastric_2004,pal_stomach_2007}.
Stomach motility was defined from in-vivo MRI data to study the rate of gastric emptying.  Kozu et al. \cite{kozu_analysis_2010,kozu_piv_2014} 
studied the mixing of gastric contents in a simplified 2D model of the terminal antrum. An axisymmetric model was also developed by Alokaily et al. \cite{alokaily_characterization_2019} to study the effect
of peristaltic wave parameters and viscosity on the retropulsive jet and recirculation of antral contents.  Others have used full 3D stomach geometries as well \cite{singh_fluid_2007,ferrua_modeling_2010}, 
in which simplified 3D geometric model of the stomach that was symmetric about the central plane was employed. In another study, 
the effect of posture on recirculation of gastric contents was investigated using an anatomically realistic geometry
\cite{imai_antral_2013}.
More recently, Ishida et al. \cite{ishida_quantification_2019} have investigated the effect of impairment of the coordination between ACW 
and the opening/closing of pylorus for meals with different viscosities. 
Li et al. \cite{li_cfd_2021} studied the effect of TAC on the flow field and incorporated acid secretion into their model to  understand pH variations inside the lumen.  They also studied the stacking of food boluses due to foods of different densities \cite{li_mixing_2021}.  A recently described model has also incorporated fluid-structure interaction (FSI) between the fluid contents and the 
gastric muscle fibers \cite{acharya_fully_2022}.


Although the interest in this field is growing, computational models of gastric function lag far behind other  disciplines such as cardiovascular and respiratory biomechanics. One key question that has 
not been addressed by the previous studies is the effect of stomach motility on the hydrolysis of  food. 
Mixing in the stomach facilitates the chemical reaction between the 
enzymes and the ingested food, which determines the composition of the contents (``chyme'') emptied out of the stomach
into the duodenum. The coupling of chemistry and fluid dynamics within the stomach is key to understanding
the sites of enzymatic activity, effect of pH gradients, and the outcomes of impairment in stomach motility.
Trusov et al. \cite{trusov_multiphase_2016}  carried out multi-species simulations of just the 
antro-duodenal region to investigate the effect of 
secretory function disorders but they did not focus on the effect of stomach motility on food breakdown.
Furthermore, since the parietal cells and the chief cells, which secrete gastric enzymes, are located in the proximal stomach \cite{bornhorst_gastric_2017},  the 
complete stomach geometry is required to accurately capture the transport and activation of enzymes.  A full stomach model that 
incorporates food hydrolysis will facilitate an understanding of the regional differences in enzymatic activity and how these differences evolve in time during gastric digestion..

The effect of stomach motility on gastric digestion can also be studied via computational models since these models allow us to systematically vary key parameters and exclude confounding effects. For example, stomach motility shows a wide variation in response to food properties but may also be affected by disease and dysfunction. Gastroparesis refers to delayed gastric emptying in the absence of any physical obstruction at the gastric outlet. In 2007, the age-adjusted prevalence of gastroparesis in the U.S. per 100,000 people was 24.2, a significant rise from 2.4 in the years 1996-2006 \cite{grover_gastroparesis_2019,camilleri_epidemiology_2011}. It can be caused by diabetes, postsurgical or postinfection complications, but it is also idiopathic in a large subset of patients
\cite{grover_gastroparesis_2019,hereijgers_antroduodenal_2022}. 
Variations in gastric motility can be caused by other conditions as well. For example, patients with Parkinson's disease have
slower gastric emptying and fewer antral contractions of smaller amplitude \cite{hardoff_gastric_2001,bornhorst_gastric_2017}. 
Erythromycin, which is a commonly used antibiotic, has been shown to improve antral contractility thereby increasing the ACW 
amplitudes \cite{annese_erythromycin_1992,parkman_gastrokinetic_1995}.
A computational model of gastric digestion could also be used to investigate on food mixing and breakdown.


In this study, we address: How do the enzymes secreted in the proximal stomach get mixed with the 
gastric contents and hydrolyze the ingested food? How does the composition of the contents emptied into the stomach
change with time? How does this mixing, hydrolysis and emptying change with variations in gastric motility?
In the past, our model has been used to study oral pill dissolution and API delivery into the intestines
\cite{seo_computational_2022,lee_computational_2022}.
Here, we use it to model the hydrolysis of a liquid meal containing protein.

\section{Methodology}

\subsection{Stomach Model}
The geometry of the stomach is obtained from the Virtual Population Library \cite{gosselin_development_2014} and is based on of the body type ``Duke'' which corresponds to a male adult. The pylorus and the duodenum are added manually using SolidWorks (Dassault Syst\`emes, V\'elizy-Villacoublay, France) because they were not resolved clearly in available dataset. The duodenum is assumed to be a tubular extension with the same diameter as the antrum. The size of the pyloric opening, which is at the junction between the antrum and the duodenum, is critical for modelling the emptying rate. In the current study, this is modelled as a circular orifice of maximum opening size 2 mm based on existing literature on the post-prandial state of the stomach \cite{meyer_human_1988,hellstrom_physiology_2006}.
\begin{figure}%
    \subfloat[Geometry and wall deformation model]{\includegraphics[width=0.49\linewidth]{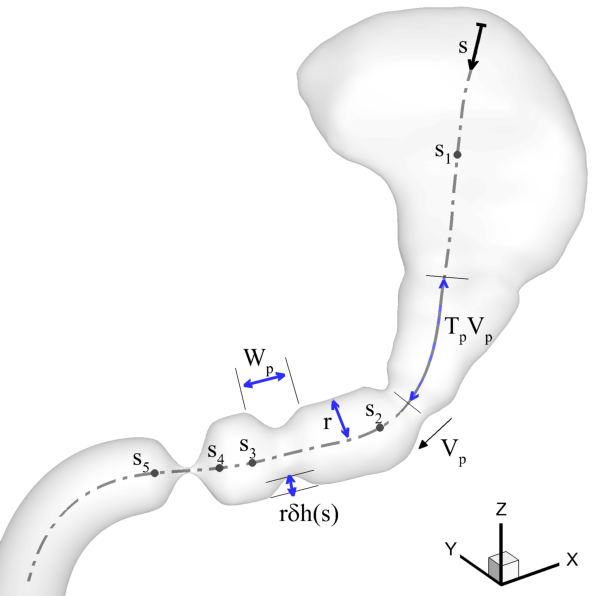}}\hfill
    \subfloat[Amplitude modulation function along the centerline]{\includegraphics[width=0.49\linewidth]{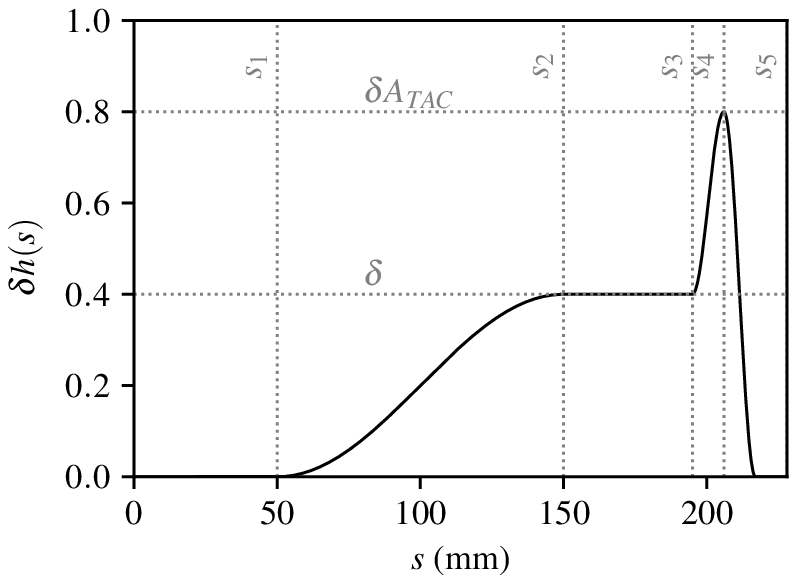}}
    \caption{\label{fig:motility} Description of the stomach geometry and the motility model. The amplitude modulation function, $h(s)$, varies the strength of the ACWs along the centerline from 0 in fundus, to $\delta$ in antrum, and to $\delta A_{TAC}$ just before the pylorus.}
\end{figure}
\begin{figure}
    \includegraphics[width=0.5\textwidth]{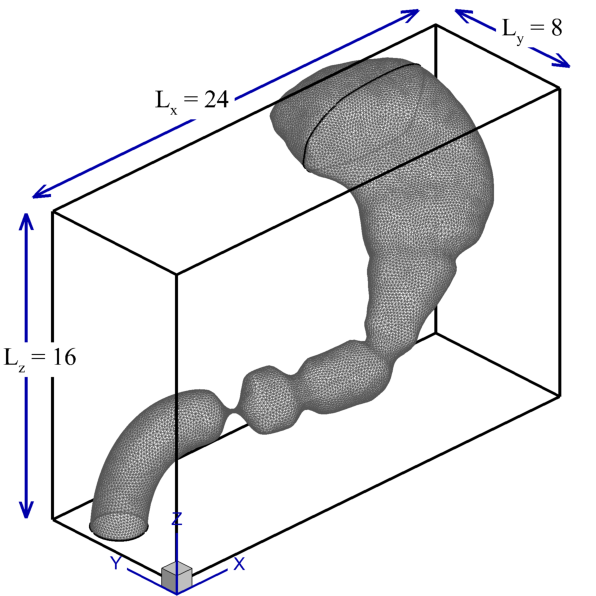}
    \caption{\label{fig:compdomain} The triangulated surface mesh is immersed in an outer Cartesian mesh with an open fundus at the top and an open duodenum at the bottom.}
\end{figure}

The amplitude of the ACW at any given axial location is given by $\delta \times h(s)$ where $\delta$
is the nominal strength of the ACWs.
The amplitude modulation function $h(s)$ is a factor used to vary the effective strength of the ACWs
from zero in the fundus (i.e. no ACW), to $\delta $
in the antrum, and then to $\delta A_{TAC}$ near the pylorus corresponding to the high amplitude collapse of walls due
to terminal antral contraction (TAC). This variation, along with the stomach geometry describing the relevant motility
parameters, is shown in Figure \ref{fig:motility}. The velocity, width and the occlusion of the ACWs were chosen from the 
literature and are described in Table \ref{tab:mparams}.
\begin{table}
\centering
\begin{tabular}{cc}\toprule
Parameter  & Value  \\\midrule
$V_p$    & 2.3 mm/s \\
$W_p$    & 20 mm    \\
$T_p$    & 20 s    \\
$\delta$ & 0.4     \\
$A_{TAC}$& 2.0     \\\bottomrule
\end{tabular}
\caption{Motility model parameters based on Ferrua and Singh\cite{ferrua_modeling_2010}. Other values of nominal amplitude of ACWs, $\delta$, 
        were also considered ($\delta = $ 0.20, 0.27, 0.34) to model the weak motility state of the stomach.}
\label{tab:mparams}
\end{table}

\subsection{Fluid Flow}
The described stomach geometry and motility model is discretized with $\sim$ 33,000 triangular elements. The
surface mesh is immersed into a Cartesian volume grid of size 24 cm $\times$ 8 cm $\times$ 16 cm in $x-$,$y-$ 
and $z-$direction, respectively, as shown in Figure \ref{fig:compdomain}, with a grid spacing of 0.5 mm in each
direction.

The gastro-duodenal contents are assumed to have the properties of water (with density $\rho = 1000$ kg/m\textsuperscript{3},
and dynamic viscosity $\mu=0.001$ Pa s) which is a good assumption for liquid meals. The fluid flow is governed by 
the incompressible Navier-Stokes equations for Newtonian fluid:
\begin{equation}\label{eq:conti}
    \nabla \cdot \vec{u}=0;
\end{equation}
\begin{equation}\label{eq:mom}
    \rho\left( \frac{\partial u}{\partial t}+\nabla \cdot \left(\vec{u}\vec{u}\right) \right)
        = -\nabla p + \mu \nabla^2 \vec{u} + \rho \vec{g},
\end{equation}
where $\vec{g}=-9.81$ m/s\textsuperscript{2} $\hat{z}$ is acceleration due to gravity.
A no-slip boundary condition is specified at the stomach lumen, and the geometry has two openings - one cutting through the 
fundus and another through the duodenum. The boundary condition (BC) on these two openings are specified based on the 
state of the pylorus. When the pylorus is open, the fundic boundary is closed by a zero-velocity BC and the duodenal
boundary acts as an outflow with a zero-gradient BC for $\vec{u}$. 
When the pylorus is closed, the roles
are switched -  a no-slip BC is specified at the duodenal boundary and the fundic boundary has a zero-gradient BC. This 
allows small amounts of inflow or outflow from the corpus as would be expected for a incompressible flow. 

The collapsing terminal antrum decreases the stomach volume when the 
pylorus is open, causing emptying of gastric contents through the pylorus and outflow via the open duodenum. The pylorus closes immediately before the completion of the collapse of terminal antrum. Subsequently, the antral relaxation raises the stomach volume back to its original value which pulls in more food from the fundus region. This flow behaviour acts as a model of tonic contractions that forces a net one-way  flow through the pylorus. The rate of emptying is determined by the ACW kinematics and the fluid properties.

These equations are solved using ViCar3D, a sharp-interface immersed boundary method (IBM) solver \cite{mittal_versatile_2008}. 
In this solver, the spatial derivatives are discretized using second-order central difference schemes, the equations are advanced in time using a fractional-step method, and the Crank-Nicolson scheme is used to integrate in time. A stabilized bi-conjugate gradient method is used to solve the pressure Poisson equation. More details on the numerical methodology and grid convergence study can be found in earlier works \cite{lee_computational_2022,seo_computational_2022}.

\subsection{Gastric Digestion Model}
The chemical kinetics in a postprandial stomach are highly complex with different nutrients undergoing different chemical processing. The breakdown of carbohydrates, for example, begins in our mouth as mastication mixes our saliva with the food bolus. This continues till salivary amylase is exposed to the acidic environment of the stomach where it is deactivated \cite{bornhorst_bolus_2012}. The digestion of carbohydrates subsequently resumes in the small intestines.  Proteins and lipids, on the other hand, are targeted by enzymes secreted from the gastric walls. The proximal gastric walls contain chief-cells that secrete pepsinogen and gastric lipase \cite{bornhorst_gastric_2017}. Pepsinogen is activated in the presence of acid to form pepsin which hydrolyzes proteins, while gastric lipase helps breakdown lipids. Each of these enzymes is highly sensitive to pH; pepsin's activity is maximum at a pH of 2, and is completely inactive  for pH$>$5; and gastric lipase on the the other hand shows peak activity in the pH range 5-6 \cite{bornhorst_gastric_2017}. The process is further complicated by other factors such as enzyme secretion rates that vary with the composition and physical properties of food as well as time elapsed since meal consumption. Thus,  several assumptions need to be made to devise a model for the chemical breakdown of a liquid meal.

For this study, we focus on a single component liquid meal that only contains protein. Lipids are not considered in this model since they are lighter than water and add the extra complexity of gravity induced buoyancy effects. We  further assume that the stomach has a constant pH at all times throughout the volume and the enzyme pepsinogen is instantaneously  activated into pepsin upon entering the  acidic environment of the stomach.

Protein hydrolysis is assumed to follow first-order catalytic reaction kinetics:
\begin{align}\label{eq:reac}
    &\text{Protein} & + \quad &\text{Pepsin}   &\rightarrow \, \,\,\text{Hydro}&\text{lyzed Protein}  &+ \quad &\text{Pepsin};\\
     &\text{(Food)} &&\text{(Enzyme)}  &\qquad        &\text{(Digesta)}         &   &\text{(Enzyme)}
\end{align}
There can be multiple varieties of proteins in a protein meal, and each type of protein exhibits a different reactivity with pepsin. The reaction constants for pepsin's catalysis of each of these proteins are however not readily available. Here we assume that the ingested meal contains only one kind of protein - N-acetyl-L-phenylalanyl-L-phenylalanine ($C_{20}H_{22}N_{2}O_{4}$) - for which the reactivity with pepsin has been studied and the  reaction constants at optimal pH value are available \cite{cornish-bowden_rate-determining_1969}.

The three species in the model - pepsin, protein, and digesta - are modelled as passive scalars and an advection-diffusion-reaction equation is used to solve for their concentration:
\begin{equation}\label{eq:ad}
    \frac{\partial c_i}{\partial t}+ \left(\vec{u} \cdot \nabla\right) c_i = D_i\nabla^2 c_i +S_i
\end{equation}
where $c_i$ is the concentration of $i^{th}$ species ($i=1,2,3$ correspond to pepsin, protein and digesta, respectively),
$D_i$ is the diffusion coefficient of that species, and $S_i$ is the source term due to the digestion reaction. Given that the mixing in the postprandial stomach is advection dominated, we assume that each 
species has the same coefficient of diffusion $10^{-9}$ m\textsuperscript{2}/s \cite{trusov_multiphase_2016}. Pepsin  acts as a catalyst an therefore $S_1 = 0$. The source term due to the digestion reaction consumes protein, and produces digesta; therefore, $S_2 = -S_3$ and following \cite{cornish-bowden_rate-determining_1969,trusov_multiphase_2016}, the value of this consumption/production rate is set to $k_1 c_2 c_1/(k_2+c_2)$, $k_1=0.038 /s$ where $c_1,c_2$ represent the concentrations of pepsin and protein respectively, and $k_2=1.40\times 10^{-3}$ kmol/$m^3$.

To discretize the advection-diffusion equation, we use a second-order central-difference scheme for the diffusion terms, and a hybrid of second-order central and upwind schemes for the advection terms. A fourth-order Runge-Kutta method is used to explicitly integrate in time.  At $t=0$, the stomach is filled with the homogeneous liquid meal, while the concentration of pepsin and digesta is set to zero. The concentration of protein in the meal is calculated based on a typical protein shake contents in the market - 24 g protein in 200 mL water. A zero mass-flux boundary condition is applied on the stomach walls for food and digesta. For pepsin, the flux on the proximal walls of the stomach is specified using the  basal-pepsin-output (BPO) reported by Feldman et al. \cite{feldman_effects_1996} . On the rest of the stomach walls, pepsin has a zero mass-flux. For  all the species, a convective boundary condition is implemented at the duodenal outflow boundary, and a zero gradient condition is applied at the fundic boundary.

\subsection{Simulating over long digestion times}

The computational grid consists of $\sim$25 million points, which were required to resolve the flow through the pylorus and, at the
same time, include the duodenum and entire stomach. Furthermore, simulating a few ACW cycles is not sufficient to capture the trends in
meal breakdown. Even though the ACWs are initiated every 20 seconds, the process of gastric digestion might take anywhere from 30 min to several hours depending on the caloric density and other meal properties \cite{koziolek_simulating_2013}. 
Each ACW cycle of 20 seconds takes $\sim$6000 hours of CPU time to simulate, and we are interested in digestion trends given by O(100) ACWs. 
This necessitates a computationally efficient approach to modelling the digestion. In the current simulations, we made use of the periodicity of the stomach kinematics and the flow inside the stomach model to enable simulation of digestion over O(100) ACWs at a considerably reduced computational expense.

We assume that the open fundus supplies the food that is emptied
in each cycle and therefore, the volume of the stomach does not change with time. This assumption is valid for the early to mid stages of the digestion of a meal. The flow inside the stomach is therefore driven by the periodic motions of the ACW peristalsis and consequently, the fluid flow in this stomach achieves a periodic state within a few ACW cycles. The L\textsubscript{2}-norm of the the difference between velocity field of the $3^{rd}$ cycle and the $4^{th}$ cycle, normalized by the velocity of the ACW, was O($10^{-7}$). In the current procedure, the simulated flow field for the 3rd cycle (roughly after about 40 seconds of ACW cycling) is recorded and subsequently, only the transport equation for pepsin, protein, and digesta are solved with this recorded velocity field. By eliminating the solving of the momentum and the mass conservation equations, we obtain a speed-up of 8.5 times and this enables us to simulate digestion for O(1000) seconds.

\section{Results}
The development of computational models of the stomach is far behind that of cardiovascular flows. Only a few studies in the past have numerically investigated gastric fluid dynamics, and in this study we model the chemo-fluid dynamics of enzymatic hydrolysis of a liquid meal. Our aim is to investigate the effect of reduced motility (due to disease or medication) on the fluid flow as well as the rate of breakdown of liquid meal. The nominal amplitude of the antral contraction waves (ACWs) is varied from 0.4 (the control case) to 0.2 (the weakest motility case). The results are discussed in three parts - fluid flow, protein hydrolysis, and quantification of advective mixing. Then we end with a discussion on future work and the limitations of our current approach.

\subsection{Fluid Flow}
\begin{figure*}
    \centering
    \includegraphics[width=0.98\textwidth]{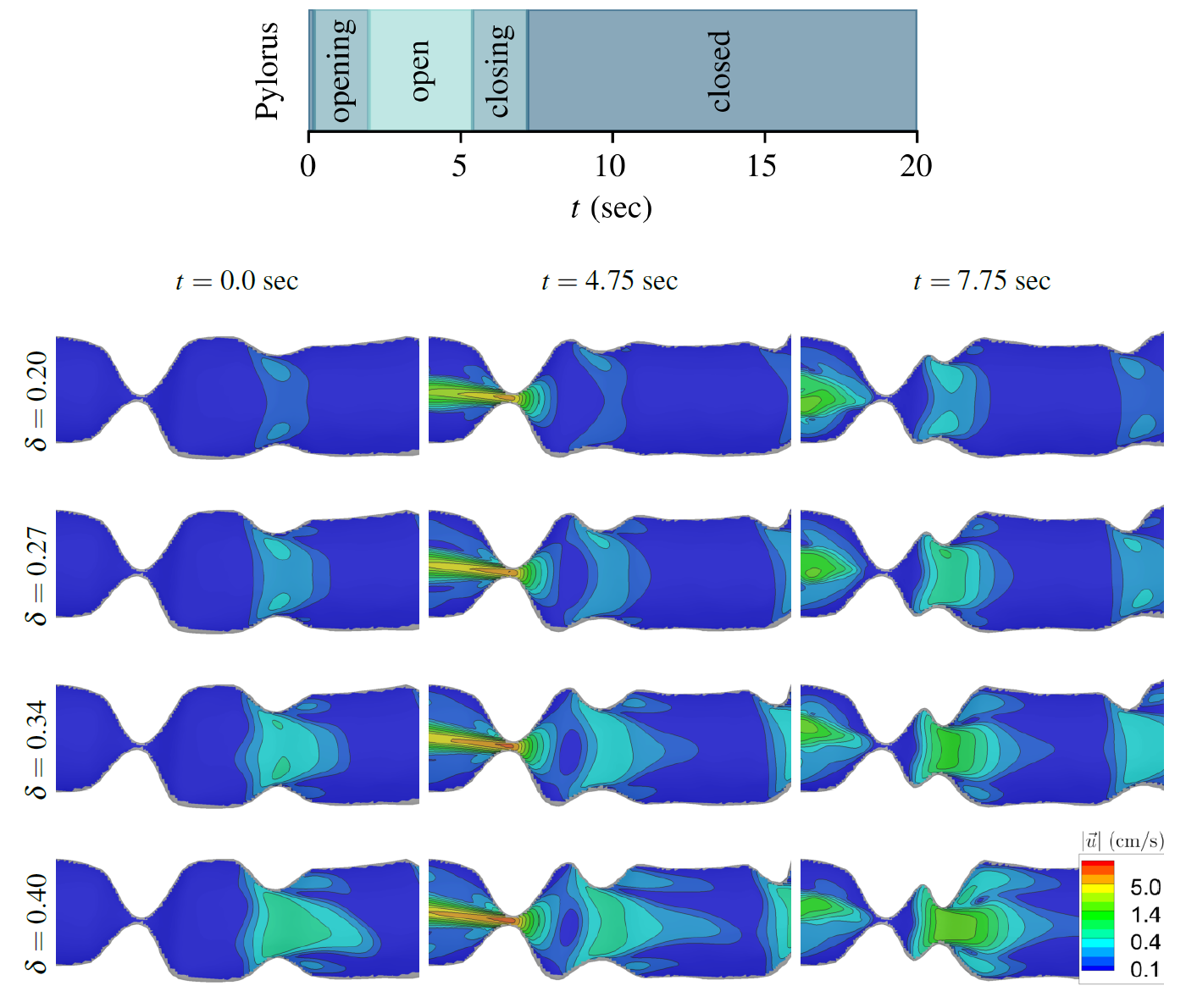}
    \caption{Comparison of velocity magnitude contours shown on a cross-section through the antro-duodenal region at three time instances - (1) $t=0.0:$ an ACW approaching an about-to-open pylorus, (2) $t=4.75:$ a fully open pylorus during the event of TAC, and (3) $t=7.75:$ right after the closing of pylorus. Notice that the TAC is still underway even after the pylorus has closed, which leads to a very prominent retrograde jet, specially in higher amplitude cases. Here $\delta$ is the fractional amplitude of the ACWs in the antrum, and the duodenum is to the left of the pylorus}
    \label{fig:umag_contours}
\end{figure*}
Previous studies have described three key features inside the stomach that plan key roles in gastric mixing and emptying- a  retropulsive jet, antral eddies between ACWs, and a pyloric jet. These flow features are reproduced in the current simulations as shown in Figure \ref{fig:umag_contours} which shows the velocity magnitude contours in the antrum on a plane that cuts through the centerline and the pyloric orifice. The retropulsive jet and the eddies between ACWs help transfer gastric contents away from the walls towards the center of the lumen and enhance mixing. The pyloric jet empties the contents into the duodenum while also entraining the pepsin into the jet, while the retropulsive jet transfers contents back into the stomach for further grinding and mixing. If the contents included solid food particles, the contents carried away by the retropulsive jet would contain all the solid particles larger than the pyloric orifice ($\sim$2 mm). This repeated grinding of food particles continues till particle size becomes smaller than the pyloric orifice, and the food can pass into the duodenum.

The time-sequence in Figure \ref{fig:umag_contours} illustrates the role of the phase between the pyloric sphincter and TAC, i.e., a collapse of the stomach walls right before the ACW terminates at the pylorus. Initially, an ACW approaches a closed pylorus and in this phase, the gastric contents are being transferred only to the proximal antrum via the retropulsive jet. As the ACW advances, the pylorus opens and a portion of the chyme trapped between the ACW and the pylorus empties via the pyloric orifice while a larger portion continues to be transported back into the stomach. This is when the TAC begins and increases the velocities of both jets. The pylorus closes before the TAC ends and the trapped chyme is, once again, left with only the retrograde path out of the collapsing segment. The retropulsive jet velocity peaks at this time while the pyloric jet velocity is maximum when the orifice was fully patent and the TAC had set in. The maximum area-averaged jet velocities of both jets are shown in Table \ref{tab:jet_vel}. The pyloric jet velocities are an order-of-magnitude higher than the retropulsive jet velocities.  Both, the retropulsive and pyloric, jet velocities also drop significantly with weakening motility. The retropulsive jet is however affected more adversely, dropping to 13\% of the velocity of the control case, as compared to pyloric jet, which dropped only to 73\%. This reveals that weakened motility implies significantly reduced gastric mixing tendencies.
    \begin{table}
        \caption{\label{tab:jet_vel} The peak velocity of the two jets is compared for different fractional amplitudes ($\delta$) of  ACWs in the antrum. The \% in parentheses is calculated by normalizing each case with the corresponding value of the control case ($\delta=0.4$). }
        \begin{tabular}{ccc}
            \toprule
            \toprule
            \multirow{2}*{$\delta$}    &       \multicolumn{2}{c}{Maximum Jet Velocity (cm/s)}\\
           \cmidrule(lr){2-3}
                        &       Retropulsive Jet    &   Pyloric Jet\\
           \midrule
            0.20        &       0.24 (13\%)                   &   7.30 (73\%)  \\
            0.27        &       0.43 (23\%)                  &   8.62 (86\%)   \\
            0.34        &       0.83 (44\%)                  &   9.43 (94\%)   \\
            0.40        &       1.88 (100\%)                  &   9.99 (100\%)  \\\bottomrule \bottomrule
        \end{tabular}
    \end{table}

The volume rate of emptying via the pyloric orifice is a key parameter because it influences the rate at which nutrients are absorbed in the intestines and it triggers the  neurohormonal feedback mechanism \cite{liu_mechanisms_2021} that controls gastric motility. The emptying rate for different ACW amplitudes are shown in  Table \ref{tab:emptrate}. The emptying rate drops by 35\% for the case with the weakest motility. 
%
    \begin{table}
        \caption{\label{tab:emptrate} Comparison of the volume of fluid emptied through the pylorus per minute for different fractional amplitudes ($\delta$) of 
                 ACWs in the antrum. The \% in parentheses is calculated by normalizing the emptying rate with that of the control
                 case ($\delta=0.4$).}
        \begin{tabular}{cc}
           \toprule
           \toprule
            $\delta$    &       Emptying Rate (mL/min)\\
           \midrule
            0.20        &       2.89 (65\%)   \\
            0.27        &       3.61 (81\%)   \\
            0.34        &       4.15 (93\%)   \\
            0.40        &       4.48 (100\%)  \\
            \bottomrule
           \bottomrule
        \end{tabular}
    \end{table}

The case with $\delta=0.4$ has the same motility parameters used by Pal et al. \cite{pal_gastric_2004} and Ferrua and Singh \cite{ferrua_modeling_2010}, which are based on MRI data obtained after ingestion of 330 mL of meal with a caloric density of 0.73 kcal/mL. This case is used to validate the model by comparing the emptying rate of $\delta=0.40$ with other studies with a meal of a similar caloric density. As seen in Table \ref{tab:validation}, the emptying rate obtained from our computational model is close to experimental  measurements by other studies of similar viscosity and caloric density meals.
%
    \begin{table}
        \caption{\label{tab:validation} The emptying rate of current model is compared with other experimental works in the literature. The meal
        properties for present work correspond to the meal that was used in the imaging study whose motility parameters 
        have been used in our model \cite{pal_gastric_2004,ferrua_modeling_2010}.}
        \begin{ruledtabular}
        \begin{tabular}{cccc}
            Study    &       Emptying Rate      &   Viscosity   &  Caloric Density \\
                     &         (mL/min)         &     (Pa s)    &       (kcal/mL)\\ 
           \hline
           Present ($\delta=0.4$)        &   4.48    &   0.001           &   0.73\\
           Marciani et al. \cite{marciani_effect_2001}  & 4.1       &   0.06            &   0.64\\
           Brener et al. \cite{brener_regulation_1983}& 4.26      & ``glucose solution''&   0.5 \\
        \end{tabular}
        \end{ruledtabular}
    \end{table}

\subsection{Protein Hydrolysis}
\newcommand\isowidth{0.19\textwidth}
\begin{sidewaysfigure}
    \setlength\tabcolsep{1pt}
    \centering
    \begin{tabular}{l c c c c c}
                    & $t=200 s$ & $t=400 s$  & $t=800 s$ & $t=2000 s$ & $t=4000 s$\\
         \rotatebox[origin=l]{90}{$\delta=0.20$} & 
         \includegraphics[width=\isowidth{}]{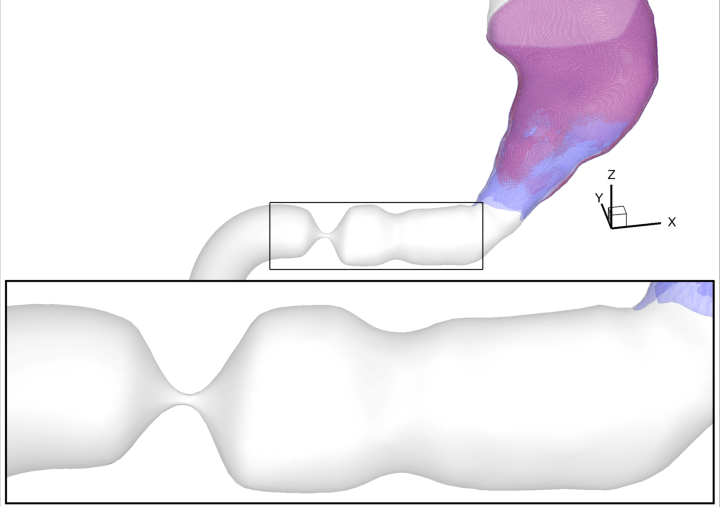} &
         \includegraphics[width=\isowidth{}]{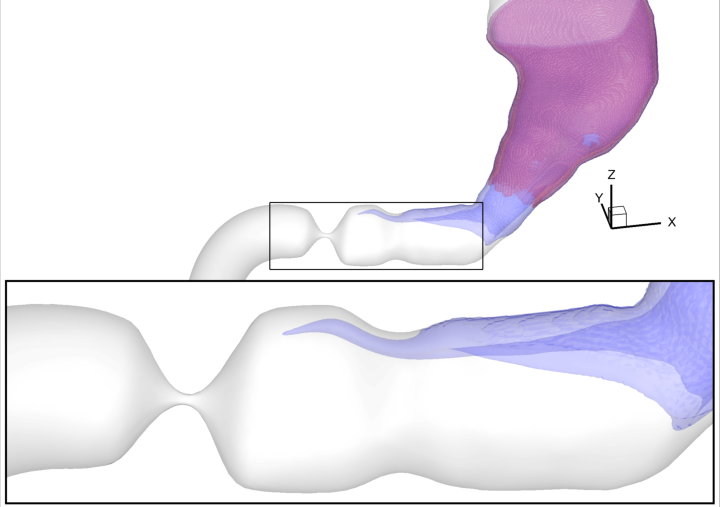} &
         \includegraphics[width=\isowidth{}]{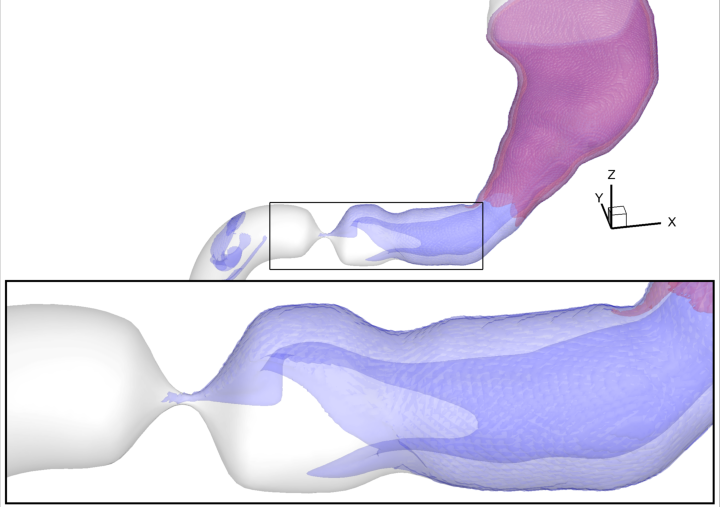} &
         \includegraphics[width=\isowidth{}]{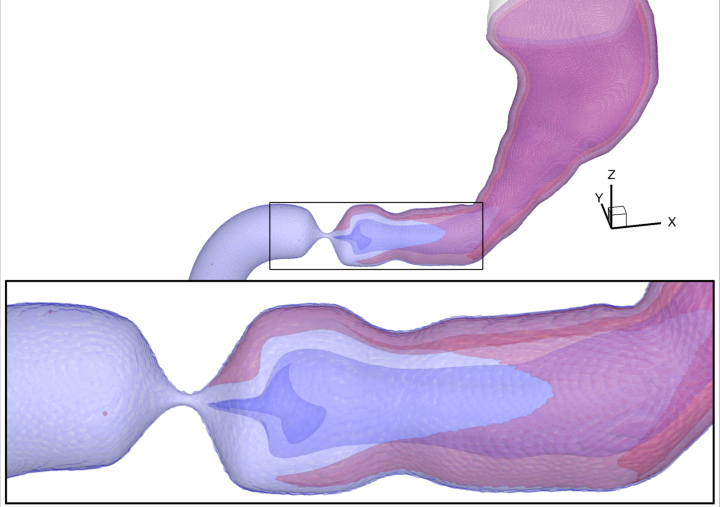} &
         \includegraphics[width=\isowidth{}]{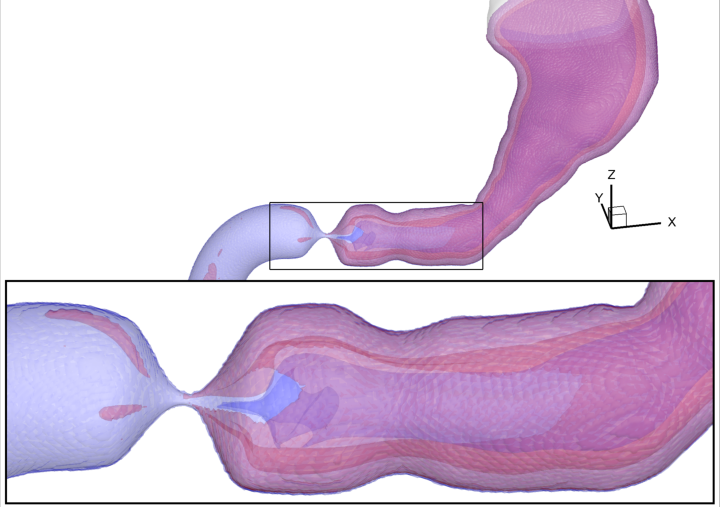} 
         \\
         \rotatebox{90}{$\delta=0.27$} &
         \includegraphics[width=\isowidth{}]{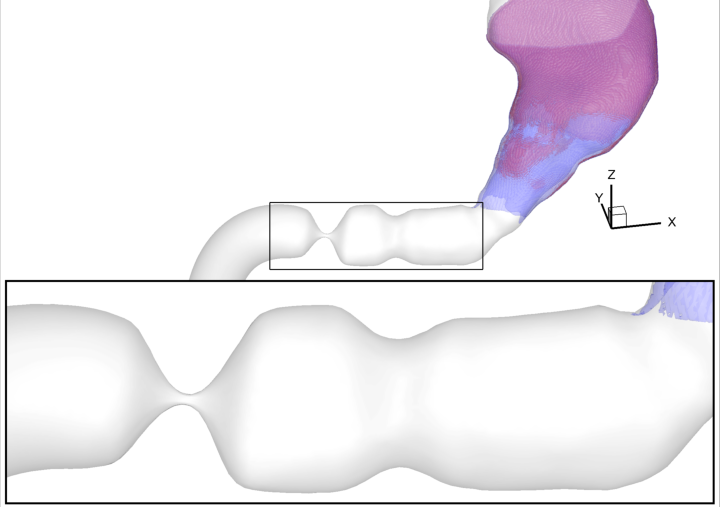} &
         \includegraphics[width=\isowidth{}]{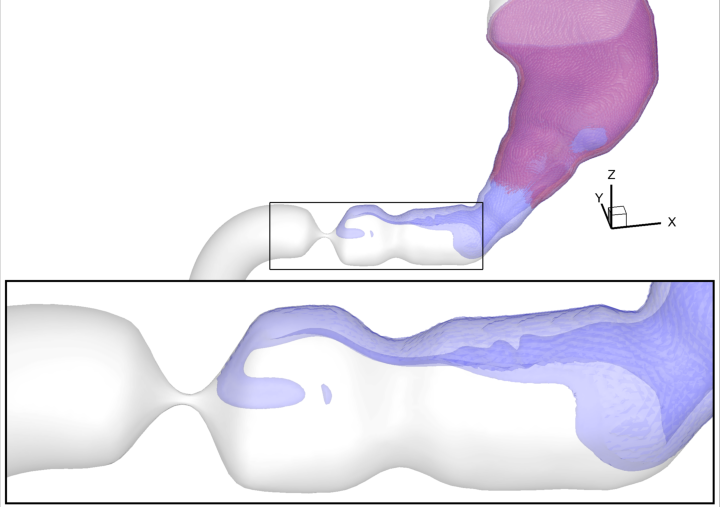} &
         \includegraphics[width=\isowidth{}]{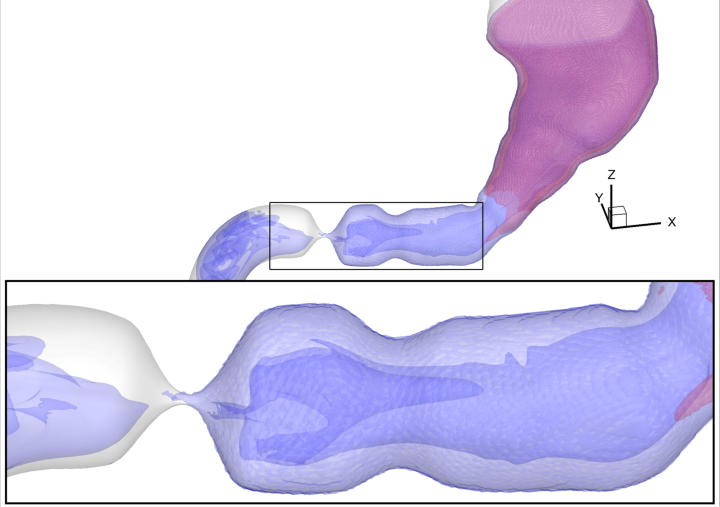} &
         \includegraphics[width=\isowidth{}]{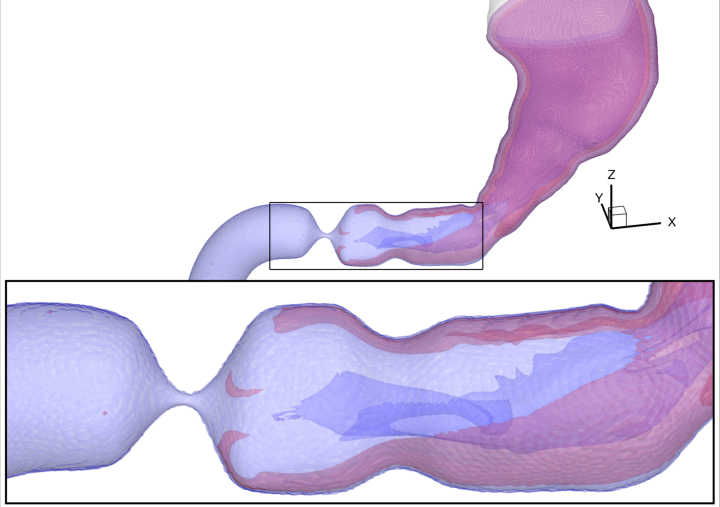} &
         \includegraphics[width=\isowidth{}]{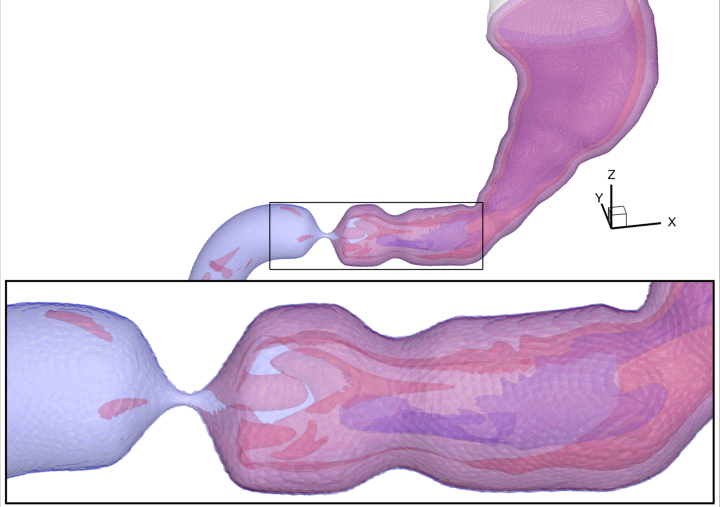}
         \\
         \rotatebox{90}{$\delta=0.34$} &
         \includegraphics[width=\isowidth{}]{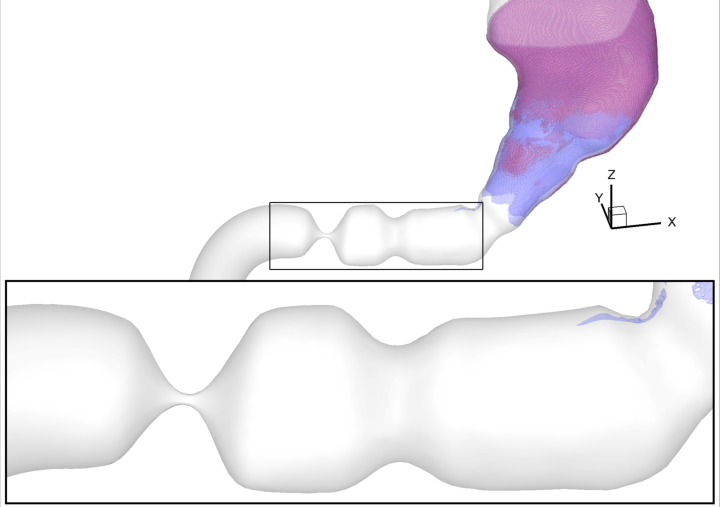} &
         \includegraphics[width=\isowidth{}]{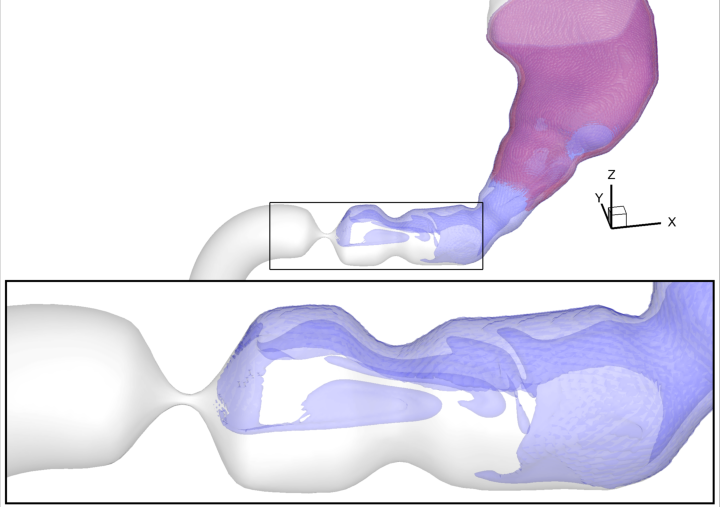} &
         \includegraphics[width=\isowidth{}]{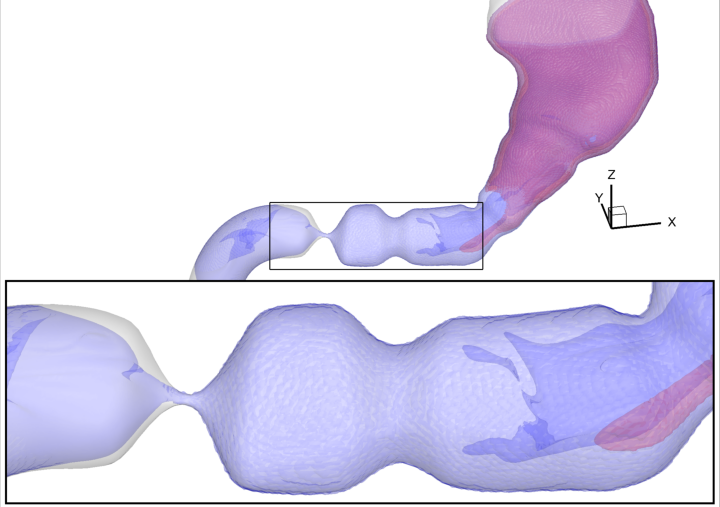} &
         \includegraphics[width=\isowidth{}]{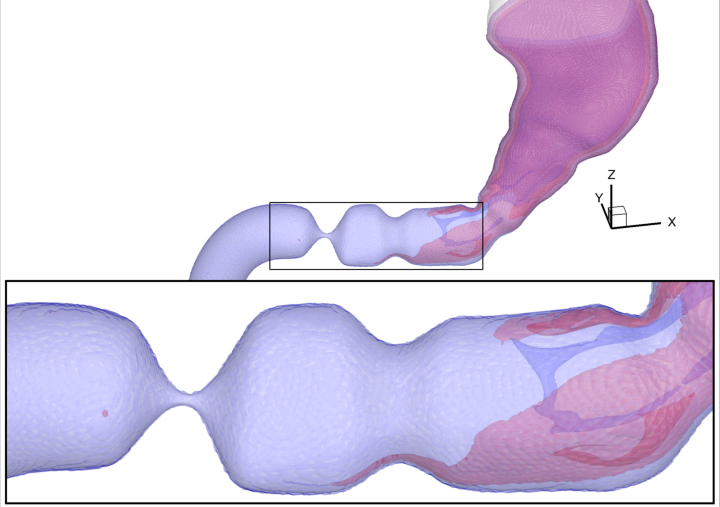} &
         \includegraphics[width=\isowidth{}]{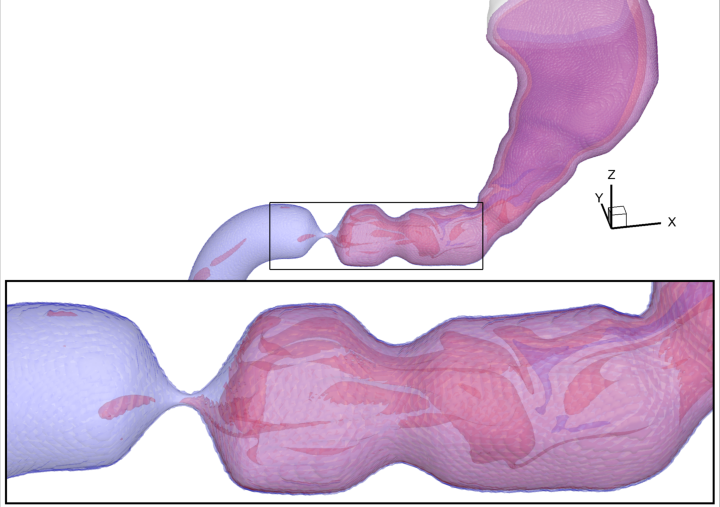} 
         \\
         \rotatebox{90}{$\delta=0.40$} &
         \includegraphics[width=\isowidth{}]{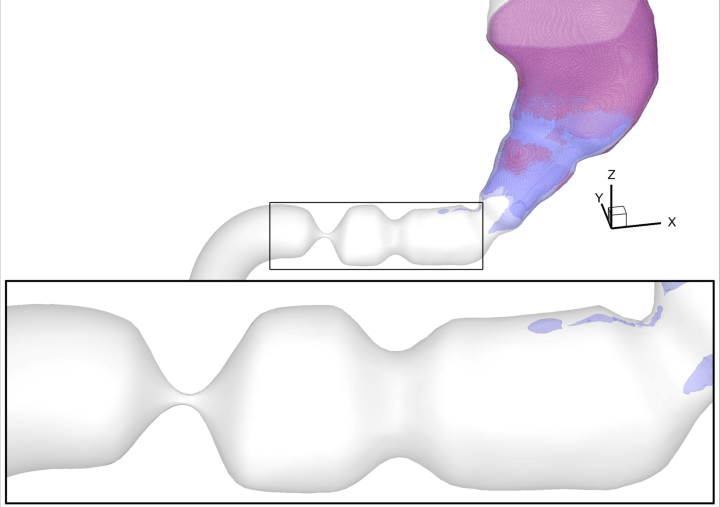} &
         \includegraphics[width=\isowidth{}]{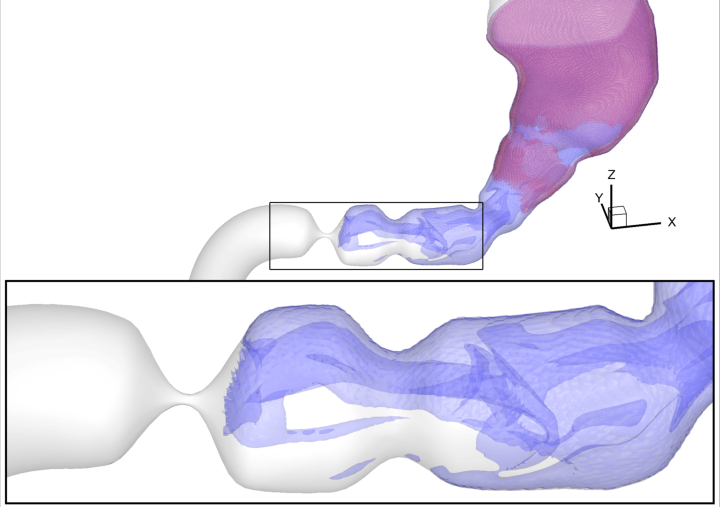} &
         \includegraphics[width=\isowidth{}]{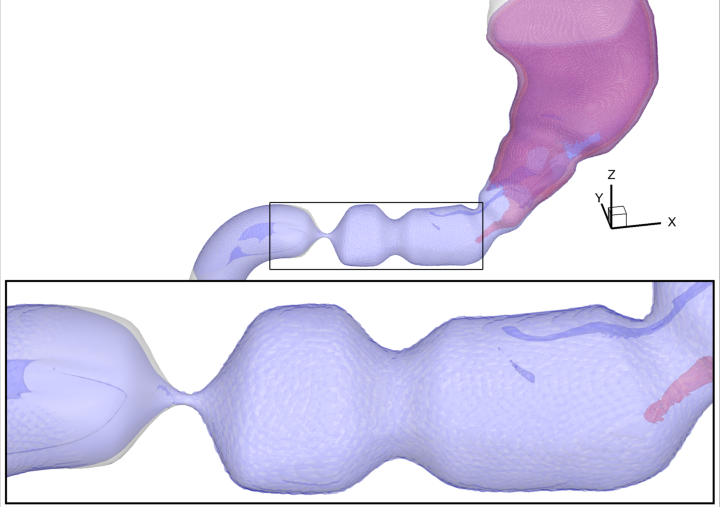} &
         \includegraphics[width=\isowidth{}]{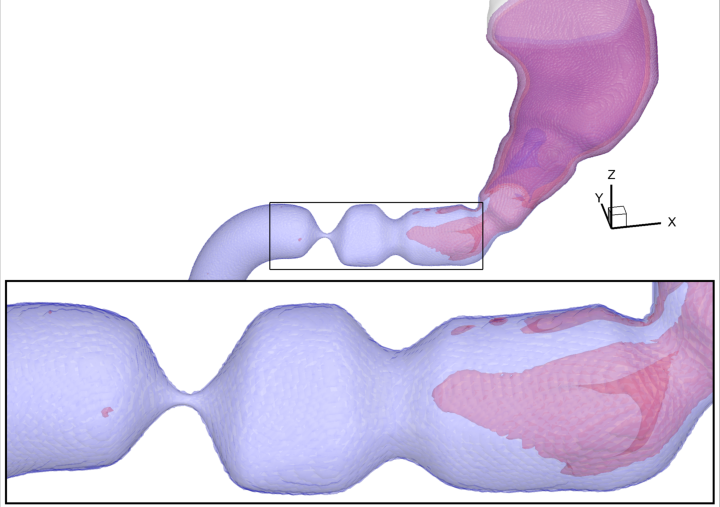} &
         \includegraphics[width=\isowidth{}]{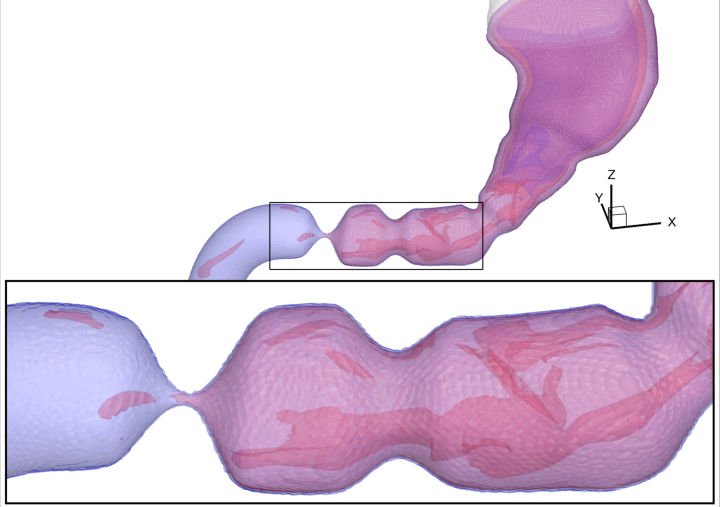}
         \\
    \end{tabular}
    \caption{Iso-surfaces of concentration at level 0.01 (normalized with respect to the initial concentration of protein in the stomach) for pepsin (red), and digesta (blue) in the antral region. The meal fills the entire stomach at $t=0$ and $\delta$
            corresponds to the fractional amplitude of the ACWs.}
    \label{fig:isosurf}
\end{sidewaysfigure}

Figure \ref{fig:isosurf} shows the progression of enzymatic hydrolysis of the protein-rich liquid meal for different amplitudes of ACW. The pepsin initiates the hydrolysis of the meal close to the secretion zones in the proximal stomach where the contents are relatively stagnant. Pepsin is subsequently transported away from the proximal walls by diffusion and also by advection in the corpus due to small amplitude ACWs that have originated in the mid-corpus. With these two mechanisms, the pepsin concentration front is slowly transported to the antrum along the stomach walls. The antrum has a much stronger advection due to the larger amplitude ACWs;  this results in a significantly enhanced mixing as compared to the proximal stomach and leads to faster hydrolysis. The weaker motility cases take longer to transport pepsin from the proximal stomach to the antrum, and are also less efficient in mixing it with the protein once pepsin arrives in the antrum. Consequently, the rate of protein hydrolysis suffers as less protein is exposed to the enzyme. A detailed analysis of this mixing is done in the next section (\ref{sec:mixing}).

\begin{figure}
    \subfloat[Protein (Unhydrolyzed)]{\includegraphics[width=0.49\linewidth]{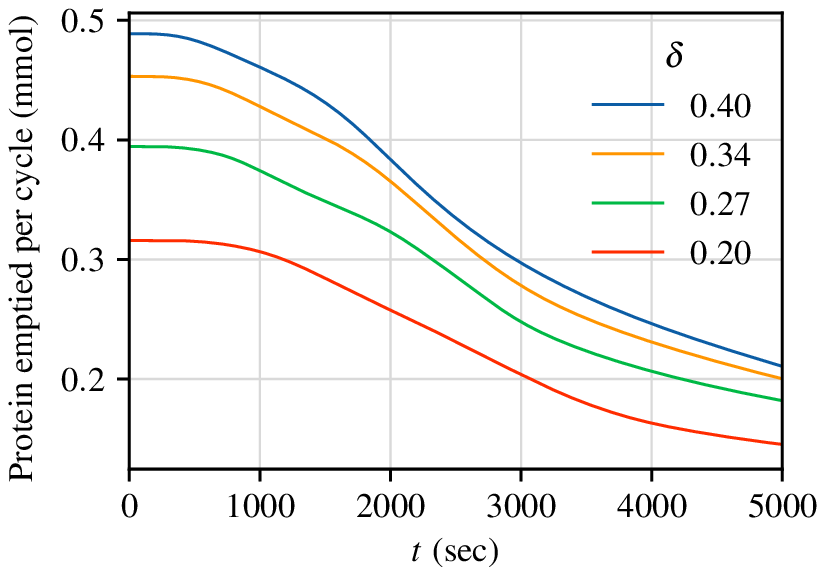}}
    \hfill
    \subfloat[Pepsin]{\includegraphics[width=0.49\linewidth]{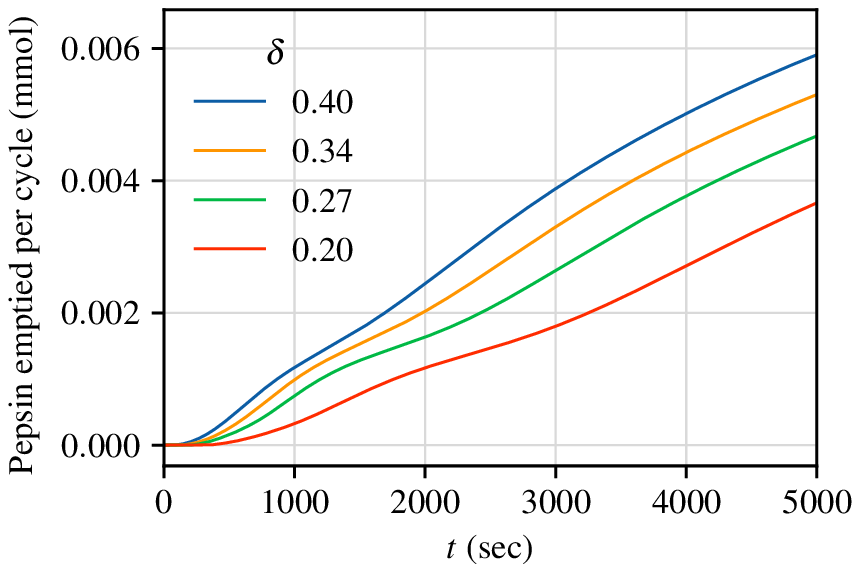}}
    \caption{\label{fig:otherflux} Amount of (a) unhydrolyzed protein and (b) pepsin emptied per cycle for different cases.}
\end{figure}
As time progresses, pepsin continues to be secreted from the proximal walls and transported towards the antrum. The increase in amount of pepsin, combined with the fact that it is not consumed during the chemical reaction, causes the rate of digestion to rise exponentially in time. Eventually, this exponential rise slows down as pepsin starts to empty via the pylorus as well (as seen in figure \ref{fig:isosurf}) and an asymptote is reached when the amount of pepsin being secreted from the walls is matched by the rate of flux of pepsin through the pylorus. The chyme that empties via the pylorus is a mixture of undigested protein, pepsin, and hydrolyzed protein. Pepsin and undigested protein fluxes are represented by Figures \ref{fig:otherflux}. In the initial stages, almost no pepsin is emptied and most protein empties without being hydrolyzed. At later times, pepsin flux rises almost linearly but the protein flux drops because mixed pepsin starts hydrolyzing at an exponential rate. As we reach $t\sim1$ hr, the exponential rate of hydrolysis starts to slow down and the protein flux starts to stagnate. During digestion, pepsin expelled into the duodenum gets deactivated because of the high pH environment in the duodenum, and from there, pancreatic proteases take over the digestion process \cite{bornhorst_gastric_2014}.

%
\begin{figure}
    \centering
    \includegraphics{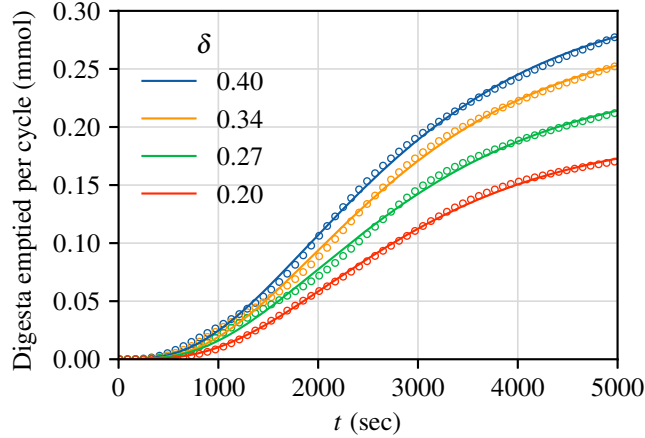}
    \caption{Amount of digesta (hydrolyzed protein) emptied per cycle for different antral contraction wave (ACW) amplitudes. The symbols denote the total amount of hydrolyzed protein emptied every 20 sec from the simulations and the curves are obtained by fitting the modified Elashoff's model to the simulation data.}
    \label{fig:digflux}
\end{figure}
\begin{figure}
    \centering
    \includegraphics{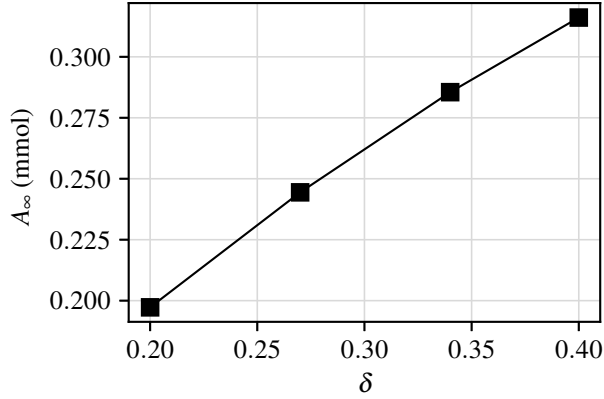}
    \caption{The amount of digesta emptied per cycle at steady state for different motility cases. This corresponds to the parameter $A_\infty$ obtained by fitting the Elashoff's model to the pyloric emptying rate obtained from simulations.}
    \label{fig:Ainf}
\end{figure}

The long duration (5000 seconds) over which the gastric hydrolysis is simulated allows us to capture the tri-phasic nature of gastric digestion as observed in Figure \ref{fig:digflux}, which shows the amount of digesta emptied per cycle through the pylorus. It also shows that the poor mixing effect in fluid flow translates to the hydrolysis of the meal as well as weaker motility cases empty lesser amounts of digesta per cycle. However, for all cases, initially there is an exponential rise which changes into a linear growth that eventually asymptotes to a value as $t\rightarrow\infty$. The figure also shows trend lines obtained by fitting  the modified Elashoff's model, which has also been used by others to capture emptying of different gastric components \cite{lee_computational_2022,kong_disintegration_2008,siegel_biphasic_1988,urbain_two-component_1989}. It is a sigmoid shape function given by:
\begin{equation}
    A(t) = A_\infty (1-e^{-\alpha t})^\beta,
\end{equation}
where $A(t)$, in the current study, corresponds to the amount of digesta emptied per cycle (20 sec), $A_\infty$ is the predicted amount of digesta that would empty per cycle at steady state,and  $\alpha$ and $\beta$ are time constants for emptying and initial delay in emptying, respectively. The parameters obtained after curve fitting are shown in Table \ref{tab:fit_params}. The parameter $A_\infty$ shows a uniform rise with $\delta$, also shown in Figure \ref{fig:Ainf}, which signifies that stronger motility leads to faster rate of hydrolysis and emptying of digesta. When the ACW amplitude is halved, the rate of hydrolysis at steady state dropped to 59\% of its original value. $\alpha$ remains roughly the same over all choices of $\delta$ which means that the time constant of these curves is similar. The delay in the emptying rate is due to the time it takes to transport pepsin from secretion zones to the antrum. The rate of secretion of pepsin was the same for all cases but weaker amplitude ACWs are expected to be slower in transporting pepsin and show a longer delay. The interface length $\beta$ first decreases with rising $\delta$, and then remains roughly the same, then decreases again - which translates to weaker motility having a large delay period with a non-linear trend.
\begin{table}
    \caption{\label{tab:fit_params}Parameters obtained by fitting the modified Elashoff's model to the digesta emptying rate in Figure \ref{fig:digflux} for different fractional amplitude of ACWs ($\delta$). Here $A_\infty$ is the predicted amount of digesta that would empty per cycle at steady state, $\alpha$ is the time constant of emptying, and $\beta$ is the initial delay in emptying, respectively.}
    \begin{ruledtabular}
        \begin{tabular}{cccc}
            $\delta$    &   $A_\infty$ (mmol) &   $\alpha$ (sec\textsuperscript{-1})    &   $\beta$\\
            \hline
            0.20        &   0.19 (59\%) &   0.00070  &   4.30\\
            0.27        &   0.24 (75\%) &   0.00068  &   3.84\\
            0.34        &   0.29 (91\%) &   0.00070  &   3.93\\
            0.40        &   0.32 (100\%)&   0.00067  &   3.55\\
        \end{tabular}
    \end{ruledtabular}
\end{table}

\subsection{Quantification of Advective Mixing}\label{sec:mixing}
\begin{figure}
    \centering
    \includegraphics[width=\textwidth]{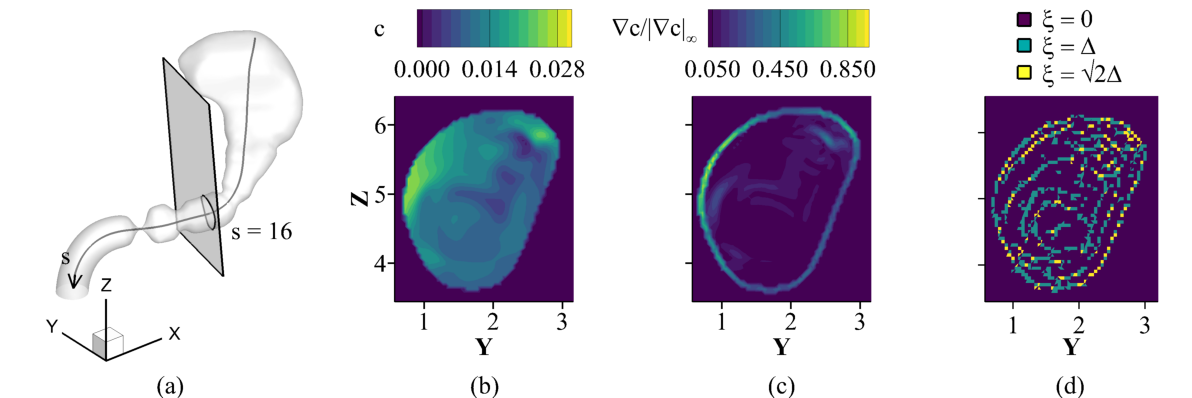}
    \caption{Calculating the interface at a cross-section through the antrum for $\delta=0.4$ case at $t=1$ hr. 
    (a) A plane normal to the centerline in the antrum ($s=16$),
    (b) contours of pepsin concentration on that plane, 
    (c) contours of normalized concentration gradient, 
    (d) the interface, $\xi$, formed by the local peaks in this gradient. }
    \label{fig:interface}
\end{figure}

\begin{figure}
    \centering
    \includegraphics[width=\textwidth]{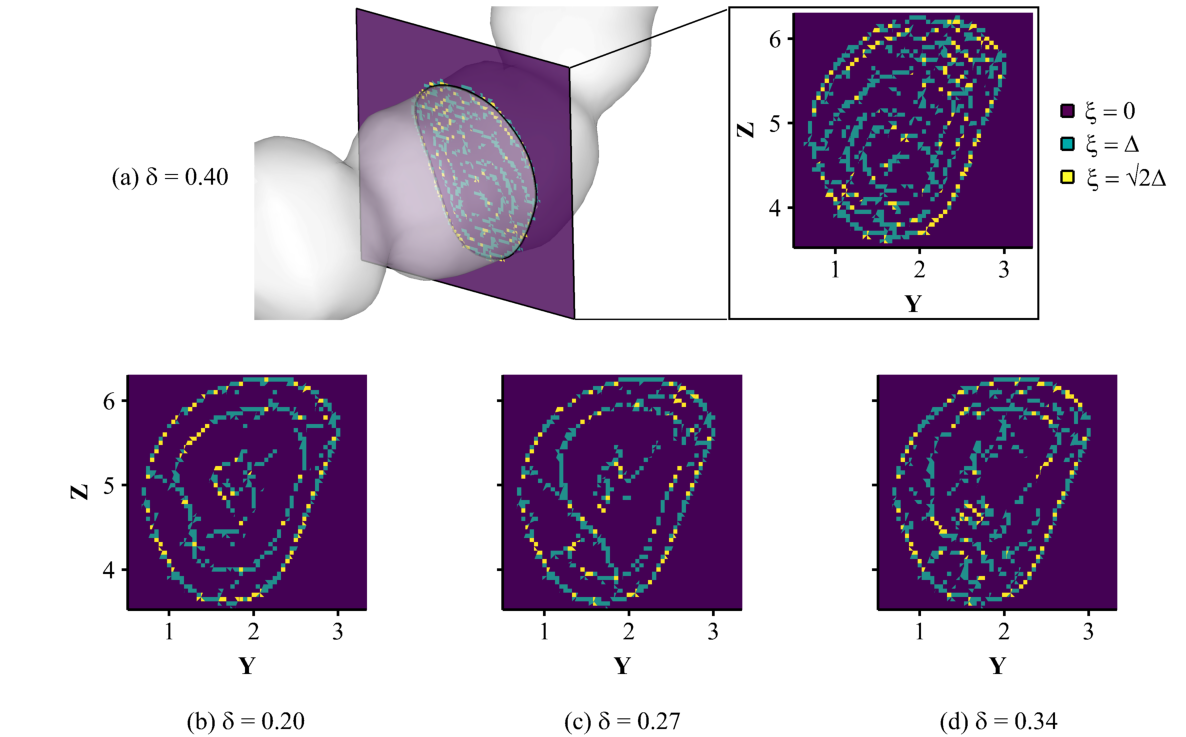}
    \caption{Reaction interfaces for different motility amplitudes ($\delta$) at $t=1$ hr at a plane perpendicular to the centerline in the antrum at $s=16$. The enhanced mixing of high amplitude cases translates to a denser distribution of reaction fronts.}
    \label{fig:ksi_t1000}
\end{figure}

The chemical reaction occurs across the interface between the food and the enzyme, i.e. protein and pepsin. ACWs promote this reaction by stretching this interface and increasing the area across which the hydrolysis can occur and many of the observations and trends regarding the hydrolysis can be explained by examining evolution of the interface between food and enzyme for the various cases. A direct measure of the interface length or area, provides a direct measure of the mixing action of the ACWs and we choose to calculate the interface for pepsin because it is the only component that is not being consumed or produced in the proteolytic process. The procedure to calculate the \emph{interface length} is mentioned in brief in the appendix (\ref{sec:app_il}), but a detailed explanation of the calculation can be found in a previous work \cite{rips_flutter-enhanced_2019}. The procedure results in an interface length field variable $\xi$ which has values of 0, $\Delta$ or $\sqrt{2} \Delta$ for each grid cell, where $\Delta$ is the linear dimension of a grid cell. This  interface length field is is visualized in Figure \ref{fig:interface} for one plane ($s=16$) at $t=1$ hr.
Figure \ref{fig:ksi_t1000} compares the interface at $x=12$ at $t=1$ hr for the four motility cases. It is clear that the
reaction interface is more widespread in the higher motility cases, which is a consequence of better mixing of pepsin with the contents. The inability to mix pepsin thoroughly results in almost non-existent reaction sites at the center of the lumen.  On the other hand, stronger motility exhibits intertwined and layered interfaces which fill the entire lumen.  

\begin{figure}
    \subfloat[]{\includegraphics[height=2.37in]{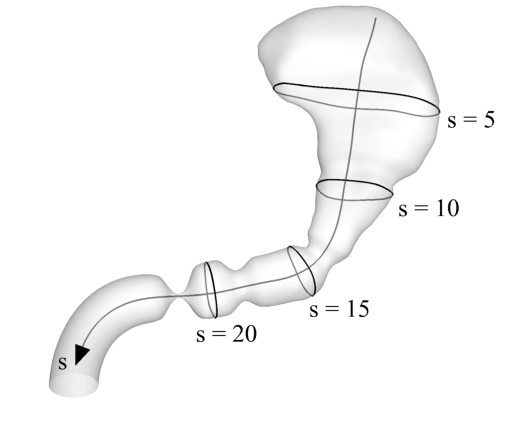}}
    \hfill
    \subfloat[]{\includegraphics[height=2.37in]{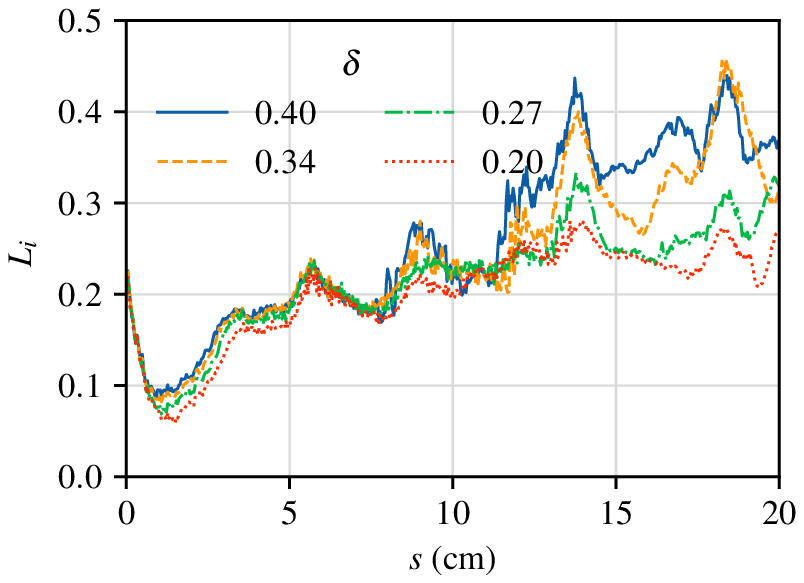}}
    \caption{The normalized area integral of $\xi$ on the cross-section planes gives interface length, $L_i$. This figure shows its comparison for different motility cases at $t=1$ hr from the fundus to the terminal antrum.}
    \label{fig:il_vs_s}
\end{figure}

The sum of $\xi$ over the cross-section plane normalized by maximum possible length of the interface at the cross-section gives us the \emph{interface length} ($L_i$), i.e.,
\begin{equation}
    L_i = \dfrac{    \sum\limits_{i,j = 1}^{N_1,N_2}       \xi(i,j)}{L_{max}} \\
\end{equation}
where, $L_{max} = N_1 \times N_2 \times \Delta$ is an upper bound on the possible interface length at any cross-section, $N_1$ and $N_2$ are the grid points inside the lumen in the two orthogonal directions, and $\Delta$ is the grid size. So, $L_i$ denotes the fraction of the lumen that is filled by the interface. The variation of the interface length along the centerline, starting from $s=0$ (at the top of the fundus) to $s=20$ (right before the pylorus), is shown in Figure \ref{fig:il_vs_s}. This figure demonstrates that all cases exhibit similar advective mixing in the fundus, but the interface lengths start to become quite different in the antrum region. This is because there is no wall motion in the fundus region, and the pepsin distribution in that region is similar for all cases.
%

\begin{figure}
    \subfloat[]{\includegraphics[width=0.49\textwidth]{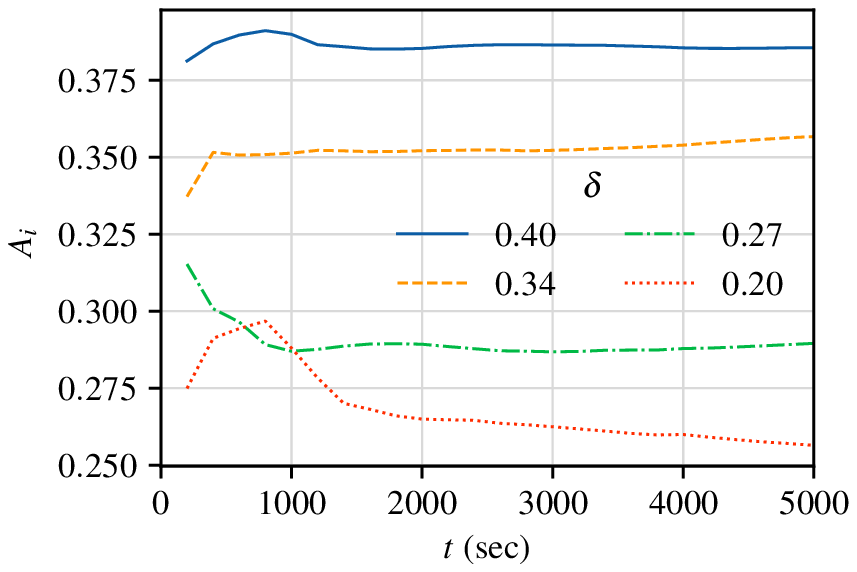}}
    \hfill
    \subfloat[]{\includegraphics[width=0.49\textwidth]{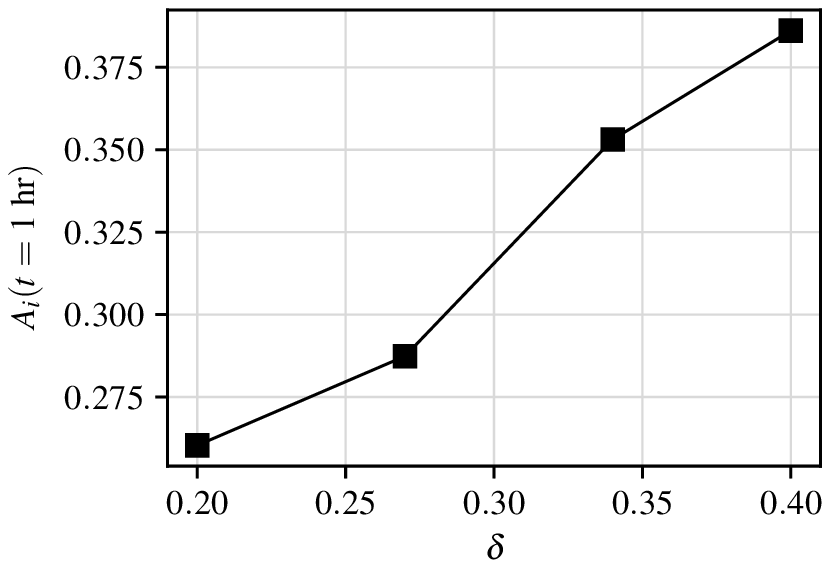}}
    \caption{(a) Interface area of pepsin is shown as a function of time. After initial transience, the interface area stabilizes to
    a constant value that quantifies how well is pepsin mixed with the gastric contents. (b) The reaction interface area at $t=1$ hr is compared for different motility cases }
    \label{fig:ia_vs_t}
\end{figure}

To see how the interface changes in time, we integrate the interface length along the centerline in the antrum region ($12\leq s\leq 20$) and normalize it by the length of the centerline to end up with a single value at each time step, termed as
\emph{interface area} ($A_i$). Mathematically,
\begin{equation}
    A_i = \frac{\int L_i ds}{\int ds}.
\end{equation}
The variation of interface area is shown in Figure \ref{fig:ia_vs_t}. This area is a measure of the size of the pepsin's concentration front and  quantifies how well it is mixed in with the contents of the stomach. There is a transience period after which all cases stabilize to a constant value. It is noted that high ACW amplitude cases not only have a shorter transience time but a significantly higher overall interface area. Furthermore, to quantify the steady state behavior, the value of interface area after an hour for each case is also shown.
The current model provides new data and insights for gastric mixing and food breakdown, but has several limitations. First, the stomach volume is held constant over the entire duration of 5000 seconds simulated here. This assumes that the fundus has sufficient meal content stored that it can keep supplying small amounts over time to maintain the stomach volume. This assumption made it computationally tractable to capture hydrolysis trends over a long enough digestion period. The model also assumed a uniform pH throughout the stomach volume and this translates to instant activation of pepsinogen into pepsin. In an actual stomach, the importance of acid secretion and pH microclimates in enzyme activation is expected to be important. We expect the digestion delay phase to last longer if we consider the effect of pH as well. 

Notwithstanding these limitations, the current model allows us to quantify the effects of disorders like gastroparesis on food hydrolysis. We have quantified advective mixing within the stomach  and how this mixing is modulated by changes in antral motility. All this information would be extremely challenging to obtain in an experimental setting. We were also able extend the simulations to long durations and capture trends over a time period that is comparable to actual digestion time scale. In future work, we plan to include the effect of meal properties and tonic contractions of the stomach in our models. These features will make this model more complete and viable for a wider range of gastric digestion scenarios.

\begin{acknowledgments}
We acknowledge research funding from the National Science Foundation (NSF) award number CBET 2019405 and the National Institutes of Health (NIH) award number 5R21GM139073–02. This work used the Extreme Science and Engineering Discovery Environment (XSEDE) \cite{towns_xsede_2014}, which is supported by the NSF grant number ACI-1548562, through allocation number TG-CTS100002.

\end{acknowledgments}
\section*{Data Availability Statement}
The data that support the findings of this study are available from the corresponding author upon reasonable request.

\appendix

\section{Stomach Motility}\label{sec:app}

Stomach motility is defined in a manner similar to the model of Ferrua and Singh \cite{ferrua_modeling_2010}. The peristaltic contractions, known as antral contraction waves 
(ACWs), deform the walls towards the centerline and propagate distally - from the upper body of the stomach towards the pylorus.  The deformation was defined as:
\begin{equation}
    \vec{x}_w = \vec{x}_{w,0}+\lambda \vec{r}_w,
\end{equation}
where $\vec{x}_{w}$ is the position vector of any point on the wall, $\vec{x}_{w,0}$ is the initial position vector of the  point without any  deformation, $\lambda$ is the fractional change in the radius, and $\vec{r}_w$ is a vector from the point on the wall towards the centerline. $\lambda$ is given by:
\begin{equation}\label{eq:lambda}
    \lambda (t,s) = \delta h(s)\sum_n \frac{F(s,s_{0,n})}{2}\left(\cos\left(2\pi \frac{s-s_{0,n}}{W_p}\right)+1\right),
\end{equation}
where $s$ is the distance along the centerline, $\delta $ corresponds to the nominal deformation due to the wave, $h(s)$ is the amplitude modulation function which is used to vary the strength of this deformation along the centerline, $n$ is the wave count, $s_{0,n}$ is the location of $n^{th}$ wave,
$F(s,s_{0,n})$ is the filter function that restricts the deformation to half-wave-width distance from the center of each wave, and $W_p$ is the width of the wave.

That amplitude modulation function and filter function are given by:
\begin{equation}
    F(s,s_{0,n}) = 
            \left\{
            	\begin{array}{ll}
            		1  & \mbox{if } |s-s_{0,n}| \leq W_p/2, \\
            		0 & \mbox{otherwise, } 
            	\end{array}
            \right.
\end{equation}
\begin{equation}
    h(s) =
            \left\{
                \begin{array}{ll}
                     0                                                      &  \mbox{if }       s\leq s_1,\\
                     \frac{1}{2}\left(1-\cos(\pi\frac{s-s_1}{s_2-s_1})\right) &  \mbox{if } s_1<  s\leq s_2,\\
                     1                                                      &  \mbox{if } s_2<  s\leq s_3,\\
                     \frac{A_{TAC}-1}{2}\left(1-\cos(\pi\frac{s-s_3}{s_4-s_3})\right) & \mbox{if } s_3<  s\leq s_4,\\
                     \frac{A_{TAC}}{2}\left(1+\cos(\pi\frac{s-s_4}{s_{pyl}-s_4})\right) & \mbox{if } s_4<  s\leq s_{pyl},\\
                     0                                                      & \mbox{if } s_{pyl} <  s,
                \end{array}
            \right.
\end{equation}
\begin{equation}
    s_{0,n} = V_pt+nT_pV_p \mbox{ and } s_{pyl} = \left(s_4+s_5\right)/2,
\end{equation}
where $s_1$ to $s_5$ are locations along the centerline, $s_{pyl}$ is the location of the pylorus, 
$A_{TAC}$ is the factor that controls the strength of collapse of the terminal antrum due to TAC, 
$V_p$ is the speed of propagation of the ACW, and $T_p$ is the time interval between two subsequent ACWs.

\section{Calculating Interface}\label{sec:app_il}

\begin{figure*}
    \includegraphics[width=\textwidth]{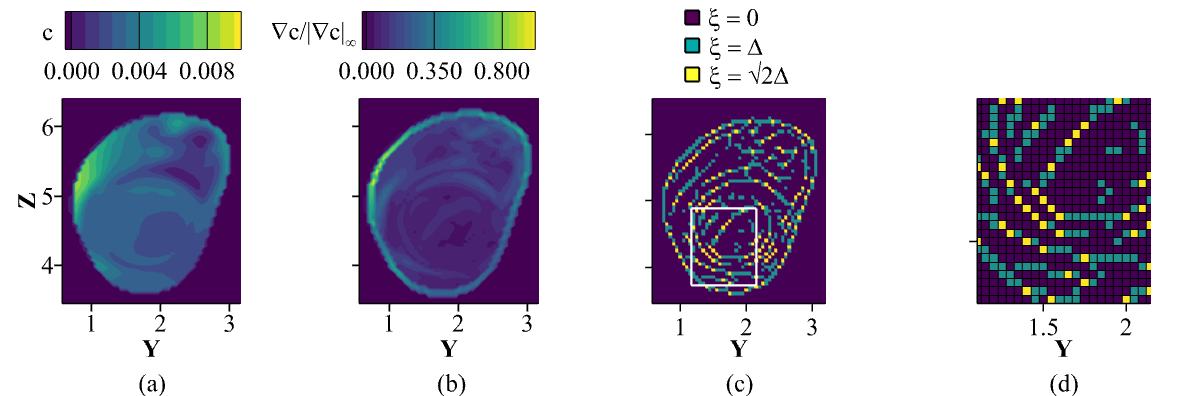}
    \caption{The procedure for calculating interface, $\xi$, at a plane perpendicular to the $yz$-plane at $x=12$. (a) The concentration of pepsin on the plane, (b) normalized concentration gradient, (c) local peaks in the gradient over the plane give the interface,  (d) zoomed into the inset with visible grid lines.}
    \label{fig:int_procedure}
\end{figure*}

 To identify the interface of pepsin, we consider planes passing normal to the stomach centerline. On each plane, we calculate the concentration gradient and then normalize it by the $L_\infty$-norm, i.e. $\nabla c/|\nabla c|_\infty$. We then find local peaks in this normalized gradient over the plane to identify the interface $\xi$. If the normalized gradient of a cell is greater than two of its immediate neighbours (north, south, east, and west), then $\xi=\Delta$. If the normalized gradient value is greater than all four of its neighbours, then $\xi=\sqrt{2}\Delta$. Each step of this procedure is visualised in Figure \ref{fig:int_procedure} at one plane at $t=1$ hr. The generated $\xi$ field inside the stomach identifies the concentration front of pepsin. Similar approach has been used in a previous work by Rips and Mittal \cite{rips_flutter-enhanced_2019}.

\bibliography{bibfile}

\end{document}